\documentclass[11pt]{article}
\usepackage{graphicx}
\usepackage{amsmath}

\title{\textbf{ ELECTRONS FROM MUON DECAY IN BOUND STATE} }

\author{\textbf{Rashid M. Djilkibaev$^{1}$ \thanks{Permanent address:
     Institute for Nuclear Research, 60-th Oct. pr. 7a,
     Moscow 117312, Russia}\ ,
     Rostislav V. Konoplich$^{1,2}$}\\
   \normalsize$^{1}$Department of Physics, New York University,
   New York, NY 10003\\
   \normalsize$^{2}$Manhattan College, Riverdale, New York, NY, 10471}

\begin{document}

\maketitle

\begin{abstract}
We present results of a study of the muon decay in orbit (DIO) contribution to 
the signal region of muon - electron conversion.
Electrons from DIO are the dominant source of background
for muon - electron conversion experiments because the endpoint of DIO electrons
is the same as the energy of electrons from elastic muon - electron conversion. 

The probability of DIO contribution to the signal region was considered
for a tracker with Gaussian resolution function and with a realistic
resolution function obtained in the application of pattern recognition
and momentum reconstruction Kalman filter based procedure to 
GEANT simulated DIO events.
It is found that the existence of non Gaussian tails in the realistic 
resolution function does not lead to a significant increase in DIO
contribution to the signal region. 

The probability of DIO contribution to the calorimeter signal was studied 
in dependence on the resolution, assuming a Gaussian resolution function
of calorimeter. In this study the geometrical acceptance 
played an important role,
suppressing DIO contribution of the intermediate range electrons
from muon decay in orbit. 

\end{abstract}

\newpage

\section*{Theoretical electron spectrum from muon decay in orbit}
\subsection*{Electron spectrum near the endpoint}

Electrons from muon decay in orbit are the dominant source of background
for muon - electron conversion experiments. In the free decay of 
a muon at rest to an electron and two neutrinos, the electron's energy
is at most half the muon rest energy, but in the decay of a bound
muon the energy approaches that of the conversion electron, $\simeq$ 105 MeV,
when the two neutrinos carry away little energy.
 
General formulas obtained in ~\cite{hanggi} and ~\cite{her}, describing the electron 
spectrum from muon decay in orbit (DIO), are complicated. 

The results of numerical calculation of the electron spectrum
for aluminum and magnesium
are presented in Table ~\ref{table:spectrum}.

\begin{table}[htb!]
\begin{center}
\begin{tabular}{|l|c|c|l|c|}
\hline
\multicolumn{5}{|c|}{$^{27}Al$ Watanabe ~\cite{wat} ~~~~~~~~~~~~~~~~~~
                       $^{nat}Mg$ Herzog ~\cite{her}}\\
\hline
Energy (MeV) & N(E) (MeV$^{-1}$)&~~~~~ &Energy (MeV)&N(E) (MeV$^{-1}$)\\
\hline
5.0     &$1.17 \cdot 10^{-3}$ &~~~~~ &5.11&$1.249 \cdot 10^{-1}$ \\
\hline
10.0     &$4.1 \cdot 10^{-3}$ &~~~~~ &10.22&$4.382 \cdot 10^{-1}$ \\
\hline
15.0     &$8.23 \cdot 10^{-3}$ &~~~~~ &15.33&$8.814 \cdot 10^{-1}$ \\
\hline
20.0     &$1.31 \cdot 10^{-2}$ &~~~~~ &20.44&1.41  \\
\hline
25.0     &$1.85 \cdot 10^{-2}$ &~~~~~ &25.55&1.981  \\
\hline
30.0     &$2.38 \cdot 10^{-2}$ &~~~~~ &30.66&2.549  \\
\hline
35.0     &$2.87 \cdot 10^{-2}$ &~~~~~ &35.77&3.068  \\
\hline
40.0     &$3.28 \cdot 10^{-2}$ &~~~~~ &40.88&3.487  \\
\hline
45.0     &$3.52 \cdot 10^{-2}$ &~~~~~ &45.99&3.706  \\
\hline
50.0     &$2.91 \cdot 10^{-2}$ &~~~~~ &51.1&2.683  \\
\hline
55.0     &$3.67 \cdot 10^{-3}$ &~~~~~ &56.21&$1.751 \cdot 10^{-1}$  \\
\hline
60.0     &$1.41 \cdot 10^{-4}$ &~~~~~ &61.32&$6.056 \cdot 10^{-3}$  \\
\hline
65.0     &$9.97 \cdot 10^{-6}$ &~~~~~ &66.43&$4.44 \cdot 10^{-4}$  \\
\hline
70.0     &$1.11 \cdot 10^{-6}$ &~~~~~ &71.54&$5.008 \cdot 10^{-5}$  \\
\hline
75.0     &$1.54 \cdot 10^{-7}$ &~~~~~ &76.65&$6.846 \cdot 10^{-6}$  \\
\hline
80.0     &$2.28 \cdot 10^{-8}$ &~~~~~ &81.76&$9.704 \cdot 10^{-7}$  \\
\hline
85.0     &$3.18 \cdot 10^{-9}$ &~~~~~ &86.87&$1.237 \cdot 10^{-7}$  \\
\hline
90.0     &$3.54 \cdot 10^{-10}$ &~~~~~ &91.98&$1.158 \cdot 10^{-8}$  \\
\hline
95.0     &$2.33 \cdot 10^{-11}$ &~~~~~ &97.09&$5.112 \cdot 10^{-10}$  \\
\hline
100.0     &$3.58 \cdot 10^{-13}$ &~~~~~ &102.2&$1.632 \cdot 10^{-12}$  \\
\hline
\end{tabular}
\caption { Decay electron spectra (DIO) for $^{27}Al$ and $^{nat}Mg$}
\label{table:spectrum}
\end{center}
\end{table}

Note that in Table ~\ref{table:spectrum} the electron spectrum
for aluminum is normalized to 1 but the electron spectrum for magnesium
is not normalized.

Near the endpoint the electron spectrum can be presented in
an analytical form found by Shanker ~\cite{shank}:

\begin{equation}
N(E) = 10^{-21} \cdot (\frac{E}{m_{\mu}})^2 \cdot
 (\frac{\delta_{1}}{m_{\mu}})^5 \cdot
  \Big (\tilde D + \tilde E(\frac{\delta_{1}}{m_{\mu}}) + \tilde F(\frac{\delta}{m_{\mu}})\Big )
\label{eqShanker}
\end{equation}

where  $\delta = E_{max} - E,~~~ \delta_{1} = E_{\mu} - E - E^{2}/2M_{A} , and $ 
E is the electron energy. The coefficients $\tilde D$, $\tilde E$, $\tilde F$ 
depending on 
nuclear charge Z, were calculated for a wide range of elements in
~\cite{shank}.

Eq.(~\ref{eqShanker}) was obtained by neglecting the variation of the
weak-interaction matrix element with energy.

The maximum possible energy of the electron is:

\begin{center}
 $E_{max} = E_{\mu} - \frac{E_{\mu}^{2}}{2 \cdot M_{A}},
 ~~~~~~~~~~E_{\mu} = m_{\mu} \cdot(1 - (\alpha \cdot Z)^{2}/2)    $
\end{center}
where $M_{A}$ is a nuclear mass, $m_{\mu}$ is a muon mass,
$\alpha$ is the fine structure constant.
In particular, $E_{max}$ is 104.963 MeV for aluminum.

In order to apply Eq.(~\ref{eqShanker}) to aluminum
the coefficients $\tilde D$, $\tilde E$, $\tilde F$ should be found for this element.
This can be done, for example, 
by fitting numerical results presented in  ~\cite{shank}
by a polynomial of the 4th power (see Appendix A).
 
As a result, for aluminum these coefficients are found to be 0.3575, 0.9483, 
2.2706, respectively.
Note that these coefficients correspond to the normalization of the
electron spectrum by the muon life time in aluminum ~\cite{measday} $\tau_{Al}=0.864 ~ \mu sec$.
The relation between the electron spectrum normalized by 
the muon life time in aluminum and the spectrum normalized to free muon 
decay is given by

\begin{center}
$N(E) = N_{free}(E) \frac{\Gamma_{free}}{\Gamma_{Al}} = 
N_{free}(E) \frac{\tau_{Al}}{\tau_{free}} \approx 0.4 \cdot N_{free}(E) $
\label{eqNorm}
\end{center}

where $\tau_{free}=2.2 ~ \mu sec$. Below N(E) will always refer to the spectrum
normalized to free muon decay.

According to Eq.(~\ref{eqShanker}) at E = 100 MeV the electron spectrum for aluminum
is $N(E) = 1.42 \times 10^{-13} ~ MeV^{-1}$. This number should be compared with 
$N(E) = 1.43 \times 10^{-13} ~ MeV^{-1}$, which is the result of 
numerical calculations of  ~\cite{wat} given in Table ~\ref{table:spectrum} 
but renormalized in order to take into account a muon life time in Al.
On can see that the difference in these numbers is less than $1\%$.
At 95 MeV this approximation underestimates the spectrum by approximately
30 $\%$.
Remind that Watanabe's spectrum in Table ~\ref{table:spectrum} was normalized to
free muon decay and calculated for energies below 100 MeV.

Above E = 100 MeV the precision of Shanker's formula, 
obtained in phase space approximation by neglecting the variation of
the matrix element with energy.
Therefore the description of the spectrum is improved as one approaches the 
endpoint,
because near the endpoint the DIO process is 
defined by the available phase space.
Note that the numerical
calculations  ~\cite{wat} properly take into account relativistic electron
wave functions and the effect of finite nuclear size on the wave functions.
Thus we conclude that above 100 MeV Shanker's approximation (~\ref{eqShanker})
gives a good description of the electron spectrum
with the precision better than $1 \%$.

It is important to emphasize that neglecting of the coefficients $\tilde E$ 
and $\tilde F$ at 
about 100 MeV leads to a significant underestimation (about $40\%$) of the electron 
spectrum  $N(E) = 1.0 \times 10^{-13} ~ MeV^{-1}$.

In order to simplify analytical and numerical calculations and analysis of 
the results we decided to use, instead of Shanker's formula, 
a simple two terms approximation to describe
the electron spectrum near the endpoint:

\begin{equation}
 N(E)  = C_0 \cdot \big (E_{max} - E\big )^5 +
C_1 \cdot \big (E_{max} - E\big )^6 
\label{eqNappr}
\end{equation}

For aluminum the coefficients $C_0$ and $C_1$ are determined by the following 
conditions: 
1) It is required that at E = 100 MeV  Eq.(~\ref{eqNappr}) should give
the same results 
$N(E) = 1.43 \cdot 10^{-13} ~ MeV^{-1}$
as Watanabe's spectrum in Table ~\ref{table:spectrum}, renormalized in order to 
take into account the muon life time in Al. 
2) Shanker's formula Eq.(~\ref{eqShanker}) is a good approximation
above 100 MeV, therefore at specific energy (we choose this energy as 104 MeV) 
near the endpoint the approximate formula (~\ref{eqNappr}) should
give the same result as Eq.(~\ref{eqShanker}).
These conditions give $C_0 = 0.3699 \times 10^{-16} \cdot MeV^{-6}$ and 
$C_1 = 0.02132 \times 10^{-16} \cdot MeV^{-7}$ for aluminum.
It was found that in this case in the range 
100 - 104.5 MeV the spectra differ by less than 0.7$\%$.

At 100 MeV the difference between Shanker's approximation (~\ref{eqShanker})
and two terms approximation (~\ref{eqNappr})
(and Watanabe's result) is about 0.7 $\%$.
At 95 MeV these approximations underestimate the spectrum by approximately
30 $\%$.
If one keeps only the leading term of Eq.(~\ref{eqNappr}) below 103.5 MeV
the difference with Shanker's approximation is about 3 - 8 $\%$

\subsection*{Expected number of DIO events}  

The expected number N of primary DIO events produced
in aluminum target during the time of the experiment 
can be calculated as

\begin{equation}
N = I_p \cdot \varepsilon_{\mu /p} \cdot \varepsilon_{gate} \cdot \varepsilon_{acc}\cdot T \cdot P
\label{eqN}
\end{equation}

In this equation P is the probability of DIO contribution to a signal region 
above the threshold energy.
By assuming that a proton flux $I_{p} = 4 \times 10^{13}/sec$,
muons stopping efficiency $\varepsilon _{\mu/p}$ 
in the Al target per primary proton is 0.25$\%$,
$\varepsilon_{acc} = 0.2$ is the setup overall acceptance, 
the time of experiment T = 10$^7$ sec, and that the measured efficiency
$\varepsilon_{gate}$ during the
650 nsec window, extending from 700 nsec to 1350 nsec after the pulse, is 50$\%$:
\begin{equation}
N = 5 \cdot 10^{17}\cdot P \cdot \varepsilon_{acc}  .
\label{eqN1}
\end{equation}

\section*{Approximated theoretical electron spectrum}

Since the calorimeter energy cutoff can be as low as 
80 MeV and the contribution from the intermediate part
of the electron spectrum can be important. 

Using the numerical results from Table ~\ref{table:spectrum} for aluminum,
the theoretical electron spectrum can be approximated by a fit.

\begin{figure}[htb!]
  \centering
  \includegraphics[width=0.7\textwidth]{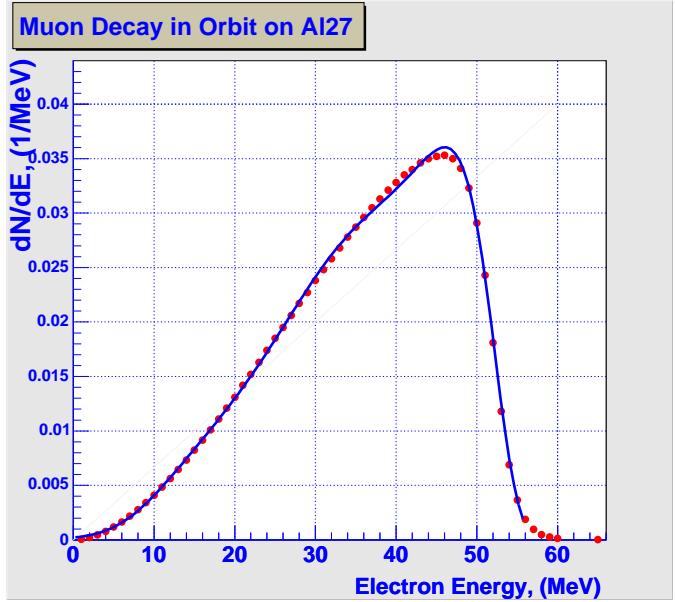} 
  \caption{
Differential electron spectrum for muon decay in orbit in Al in a linear scale.
Points are the results of numerical calculations. Solid line is a fit of 
the spectrum below 55 MeV. This spectrum
is normalized to free muon decay.
 }
\label{fig:low_spectrum}
\end{figure}

Figure ~\ref{fig:low_spectrum} shows the differential electron spectrum 
for muon decay in orbit in Al in linear scale. Points are the results of 
numerical calculations ~\cite{wat}. The solid line is an 8 parameter fit of 
the spectrum below 55 MeV in the form $e^{f(E)}$ where f(E) is a polynomial
of power 7. The parameters of the fit are presented in Appendix B. 

\begin{figure}[htb!]
  \centering
   \includegraphics[width=0.7\textwidth]{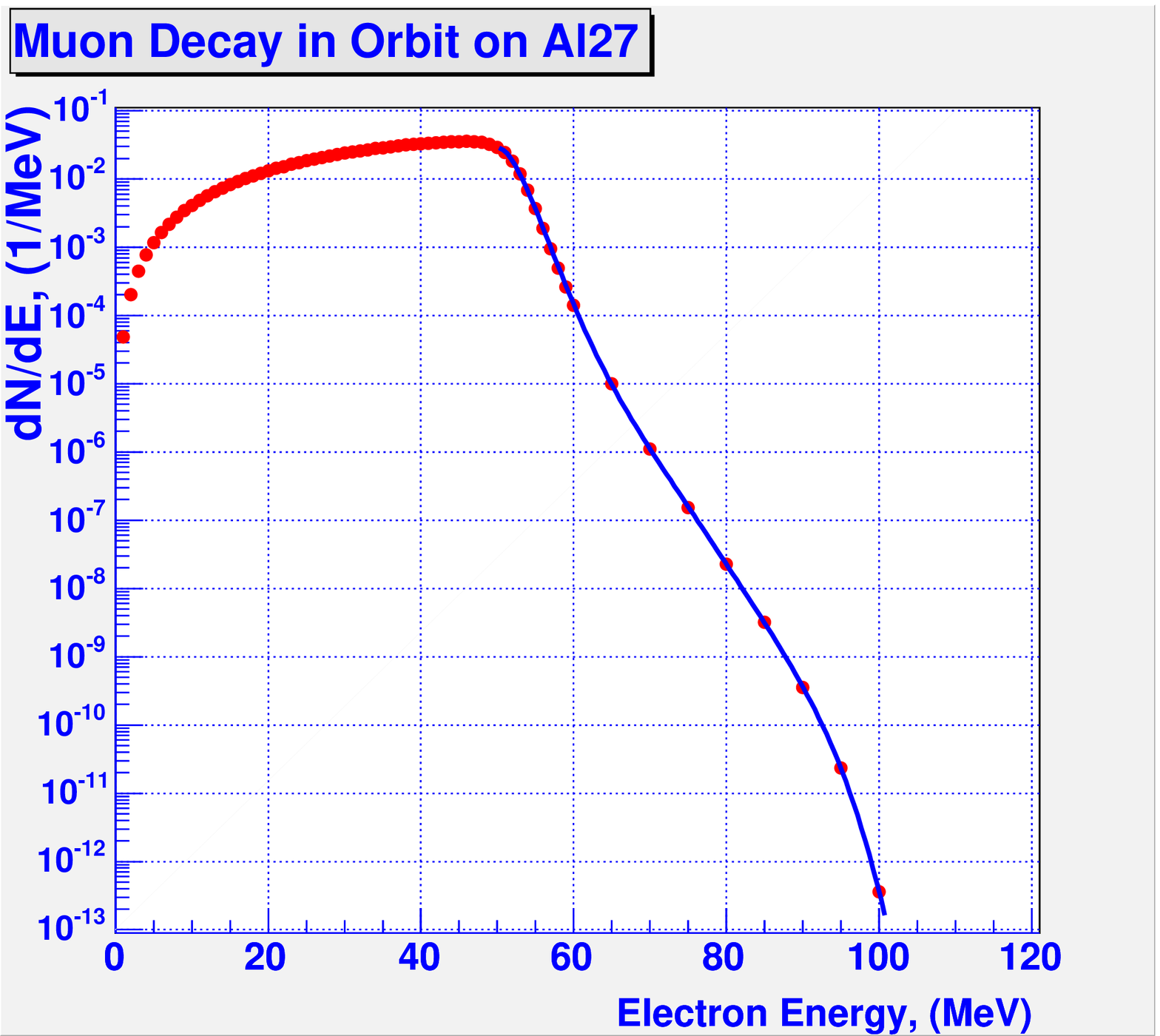}
  \caption{
Differential electron spectrum for muon decay in orbit in Al in logarithmic scale.
Points are the results of numerical calculations. Solid line is a fit of 
the spectrum in the range 55 - 100 MeV. This spectrum
is normalized to free muon decay.
 }
\label{fig:high_spectrum}
\end{figure}

Figure ~\ref{fig:high_spectrum} shows the differential electron spectrum 
for muon decay in orbit in Al,a linear scale. The points are the results of 
numerical calculations ~\cite{wat}. The solid line is an 8 parameter fit of 
the spectrum in the range 55 - 100 MeV in the form $e^{g(E)}$ where 
g(E) is a polynomial
of power 7. The parameters of the fit are presented in Appendix B. 

The region above 100 MeV is described by the two term approximation 
Eq.(~\ref{eqNappr}) but its contribution in the case of low threshold energies
is negligible.

The probability 
to produce an electron of energy above $E_{min}$  
in muon decay in orbit in Al on the basis of the approximate
theoretical electron spectrum is given by

\begin{equation}
P = \int\limits_{E_{min}}^{E_{max}} N(E) dE .
\label{int_spectr}
\end{equation}

In this expression as it was mentioned above $E_{max} = 104.963 ~ MeV$,
E is the true energy of electron (at this energy an electron is created in
muon decay in orbit).

\begin{figure}[htb!]
  \centering
   \includegraphics[width=0.7\textwidth]{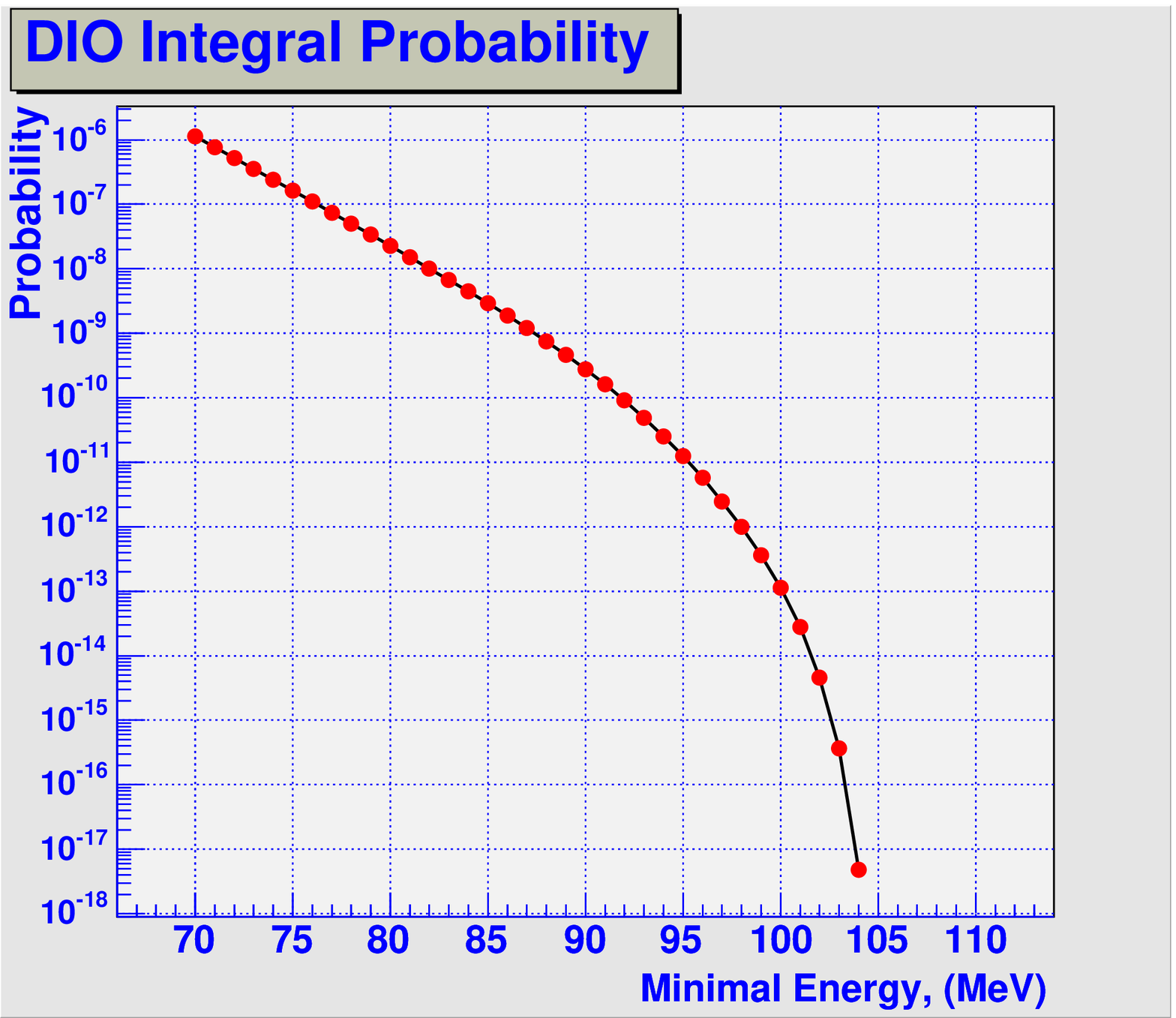} 
  \caption{
 Probability to produce an electron
of energy above $E_{min}$  
in muon decay in orbit in Al
 }
\label{fig:integral_dio}
\end{figure}

Figure ~\ref{fig:integral_dio} shows the
probability to produce an electron
of energy above $E_{min}$ in Al versus $E_{min}$.

In the range 55 - 100 MeV the approximation discussed above was used,
above 100 MeV the electron spectrum was taken in the form 
(~\ref{eqNappr}).

Using the probability, which is given by Eq.(~\ref{int_spectr}),
one can calculate the number of primary DIO
events during the time of the experiment.
It is important to emphasize that in this section we calculate the total
number of DIO electrons produced in aluminum target and we do not need
to include in this calculation a resolution function and overall 
acceptance.   

The number of DIO events, which is calculated using Eq.(\ref{eqN}) with 
$\varepsilon_{acc} = 1$,
is sharply increasing with decreasing $E_{min}$.
In particular , for $E_{min}$ = 100 MeV - $N = 5.7 \times 10^4$,
for $E_{min}$ = 95 MeV - $N = 6.2 \times 10^6$, and 
for $E_{min}$ = 90 MeV - $N = 1.39 \times 10^8$.

Table ~\ref{table:total_dio} shows the probability P and the 
number of DIO events N with an electron energy in the given 
energy range.  

\begin{table}[htb!]
\footnotesize
\begin{center}
\begin{tabular}{|c|c|c|c|c|c|}
\hline
& & & & & \\
Range &80 - 85 &85 - 90 &90 - 95 &95 - 100 & $> 100$\\
& & & & & \\
\hline
& & & & & \\
Prob.     &$1.97 \times 10^{-8}$ &$2.62 \times 10^{-9}$ &$2.64 \times 10^{-10}$ &$1.23 \times 10^{-11}$ &$1.15 \times 10^{-13}$ \\
& & & & & \\
\hline
& & & & & \\
N     &$9.85 \times 10^9$ &$1.31 \times 10^9$ & $1.32 \times 10^8$&$6.15 \times 10^6$ &$5.7 \times 10^4$ \\
& & & & & \\
\hline
& & & & & \\
A &$9.12 \times 10^{-9}$ &$1.272 \times 10^{-9}$ &$1.416 \times 10^{-10}$ &$9.32 \times 10^{-12}$ &-- \\
& & & & & \\
\hline
& & & & & \\
B &2.53823 &2.27755 &1.83767 &1.19741 &-- \\
& & & & & \\
\hline
\end{tabular}
\caption { Expected number N of primary electrons produced in muon decay in orbit for different energy ranges (in MeV). Coefficients A and B of Eq.(\ref{fit_NE}) are given.}
\label{table:total_dio}
\end{center}
\end{table}

Also in Table ~\ref{table:total_dio}, the coefficients A and B are given,
which are necessary for Monte Carlo simulation of DIO events. For this
simulation it is assumed that each 5 MeV interval from $E_1$ to $E_2$ 
of the electron spectrum in the range 55 - 100 MeV can be approximated by 

\begin{equation}
N(E) = A \cdot e^{-(E - E_1)/B}
\label{fit_NE}
\end{equation}

It is important to note that in order to simulate a process with 
the threshold in measured energy $E_m^{th}$ one has to take 
into account DIO electrons produced in target starting 
approximately at $E^{th} = E_m^{th} - 2\sigma$. In particular, 
for $E_m^{th} = 80 MeV$ and $\sigma = 5 MeV$ it would correspond
to $E^{th}$= 70 MeV. According to Figure ~\ref{fig:integral_dio}
and Eq.(\ref{eqN}) in this case the number of primary DIO events would reach 
$N = 5.65 \times 10^{11}$ making event by event simulation
unfeasible. We assume that present realistic threshold for calorimeter
measured energy is about 90 MeV and the minimal true energy of DIO
electrons is about 80 MeV. This corresponds to the number of DIO
events $N = 1.13 \times 10^{10}$ which is about 50 times less than
in the case of 80 MeV threshold in measured energy.  

\section*{DIO events in a tracker}

\subsection*{Probability of DIO contribution}

The probability to have a signal from muon decay in orbit above a threshold 
$E_{max} + \Delta$ is given by

\begin{equation}
P = \int\limits_{E_{max} + \Delta}^\infty dE_M \int\limits_0^{E_{max}} N(E) \cdot f(E_M - E) dE
\label{eqP}
\end{equation}

where $E_M$ is the measured electron energy, E is the true energy at which an electron was emitted, 
$\Delta$ is the threshold energy measured from the endpoint,
N(E) is the electron spectrum of muon decay in orbit, and f is the resolution function of the tracker.

For an ideal tracker the resolution function is the delta function and in this limit the 
probability P is given by
 
\begin{center}
$P = \int\limits_{E_{max} + \Delta}^{E_{max}} N(E_M) dE_M $
\end{center}

For the high energy part of the electron spectrum, $N(E)$ can be approximated by Eq.(\ref{eqNappr}),
leading to the probability 

\begin{equation}
P = C_0 \frac{\Delta^6}{6} - C_1 \frac{\Delta^7}{7}
\label{eqIdeal}
\end{equation} 

in the case of the ideal detector.
For the threshold energy $\Delta = -0.7 ~ MeV$, according to Eq.(\ref{eqIdeal}), the probability 
$P \approx 6.9 \cdot 10^{-19}$. 
This equation is approximately valid
for the range $-5 ~ MeV \le \Delta \le 0$.   
The lower limit is defined by the validity of the two term approximation (~\ref{eqNappr}) 
for the electron spectrum.
For positive $\Delta$ the probability P is 0 for the ideal tracker. 

\begin{figure}[htb!]
  \centering
   \includegraphics[width=0.7\textwidth]{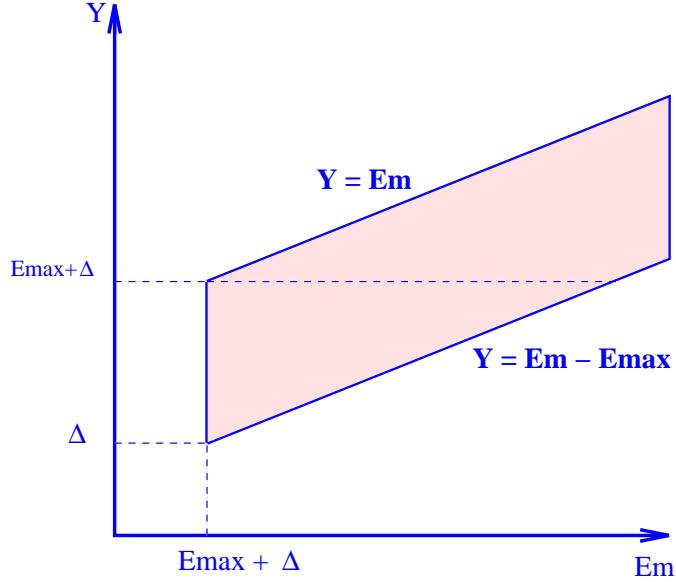}
  \caption{
  Domain of integration in $y, E_M$ variables.
 }
\label{fig:integral}
\end{figure}

To deal with a realistic resolution function it is convenient to introduce a new variable 
$y = E_M - E$ in Eq.(\ref{eqP}). The region of integration in $y, E_M$ variables is given in 
Figure ~\ref{fig:integral}. 

By interchanging the order of integration in $E_M, y$ and substituting 
$E_M = y + E$ the probability
to have a signal from muon decay in orbit above a threshold takes the form 

\begin{equation}
P = \int\limits_{\Delta}^{E_{max} + \Delta} f(y) dy \int\limits_{E_{max} + \Delta - y}^{E_{max}} N(E)  dE 
\label{eqPfin}
\end{equation}

A second integral over the region in y above $E_{max} + \Delta$ was neglected because the resolution 
function at such y is expected to be extremely small.

It follows immediately from this representation that the probability P is defined by the resolution
function above the threshold energy $\Delta$.

\subsection*{Gaussian detector response function}

Let's assume that the detector response function f(y) is of the Gaussian form:

\begin{equation}
f_G(y) = \frac{1}{\sqrt{2\pi}\cdot \sigma} ~~exp(~ -\frac{y^2}{2\sigma^2}~)
\label{eqGauss}
\end{equation}

\begin{figure}[htb!]
  \centering
   \includegraphics[width=0.7\textwidth]{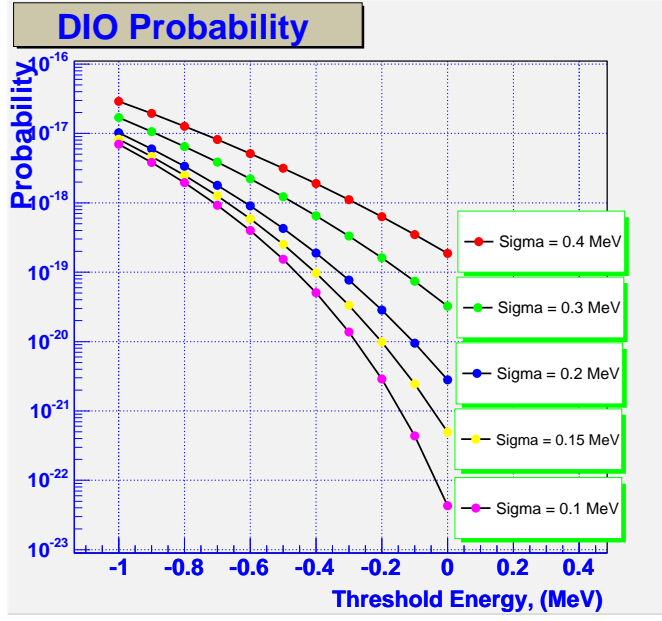}
  \caption{
    Probability of DIO contribution to a signal region as a function of 
the threshold energy $\Delta$ for different
resolutions $\sigma$.
 }
\label{fig:dio_d_s}
\end{figure}

In Figure  ~\ref{fig:dio_d_s} the probability P of muon decay in orbit as a function of
threshold energy $\Delta$ is shown for different resolutions $\sigma$.

The two term approximation for the electron spectrum Eq.(\ref{eqNappr}) was used for this plot.
For small threshold energies $\Delta \approx 0$ the probability is sensitive to the 
resolution $\sigma$ and it is proportional to $\sigma^6$.   
At $\Delta = $ -1 MeV the probabilities for $\sigma = $ 0.4 MeV and $\sigma = $ 0.1 MeV
differ by a factor 4.

Note that Eq.(\ref{eqIdeal}) defines the limiting probability
for Figure  ~\ref{fig:dio_d_s}
as the resolution $\sigma$ tends to 0. 
For the threshold energy $\Delta = -0.7 ~ MeV$ and the 
resolution $\sigma$ = 0.2 MeV the probability 
$P \approx 2.1 \cdot 10^{-18}$ is about three times greater than the probability 
$P \approx 6.9 \cdot 10^{-19}$ in the case of the ideal detector. 

If the threshold energy $\Delta > -5 ~ MeV$ the probability P,  
given by Eq.(\ref{eqPfin}), can 
be calculated analytically by using the two term approximation Eq.(\ref{eqNappr})
for the electron spectrum. In this case

\begin{equation}
P = C_0 \frac{\sigma^6}{6\sqrt{2 \pi}} I_0 + C_1 \frac{\sigma^7}{7\sqrt{2 \pi}} I_1
\label{eqPerf}
\end{equation}

The coefficients $I_0, I_1$ are given by 

\begin{equation}
I_0 = \sqrt{\frac{\pi}{2}}(15 + 45u^2 + 15u^4 + u^6)Erfc(\frac{u}{\sqrt{2}}) -
u(33 + 14u^2 + u^4)e^{-u^2/2} ,
\end{equation}

\begin{equation}
I_1 = -\sqrt{\frac{\pi}{2}}u(105 + 105u^2 + 21u^4 + u^6)Erfc(\frac{u}{\sqrt{2}}) +
(48 + 87u^2 + 20u^4 +u^6)e^{-u^2/2} ,
\end{equation}

where $u = \Delta/\sigma$, $Erfc(z) = 1 - \frac{2}{\sqrt{\pi}}\int\limits_0^z e^{-t^2} dt $

In the case of interest, $\Delta < -\sigma $, Eq.(\ref{eqPerf}) can be approximated 
by a form convenient for analysis

\begin{equation}
P = C_0 \frac{\sigma^6}{6} (15 + 45u^2 + 15u^4 + u^6) - C_1 \frac{\sigma^7}{7} u(105 + 105u^2 + 21u^4 + u^6)
\label{eqAppr}
\end{equation}

The precision of this approximation is better than $1\%$ if $\Delta < -\sigma $. 
In the limit $|\Delta| >> \sigma$ for negative $\Delta$ 
this equation reproduces Eq.(\ref{eqIdeal}).

\subsection*{Simulated detector response function}

\begin{figure}[htb!]
  \centering{\hbox{
  \includegraphics[width=0.5\textwidth]{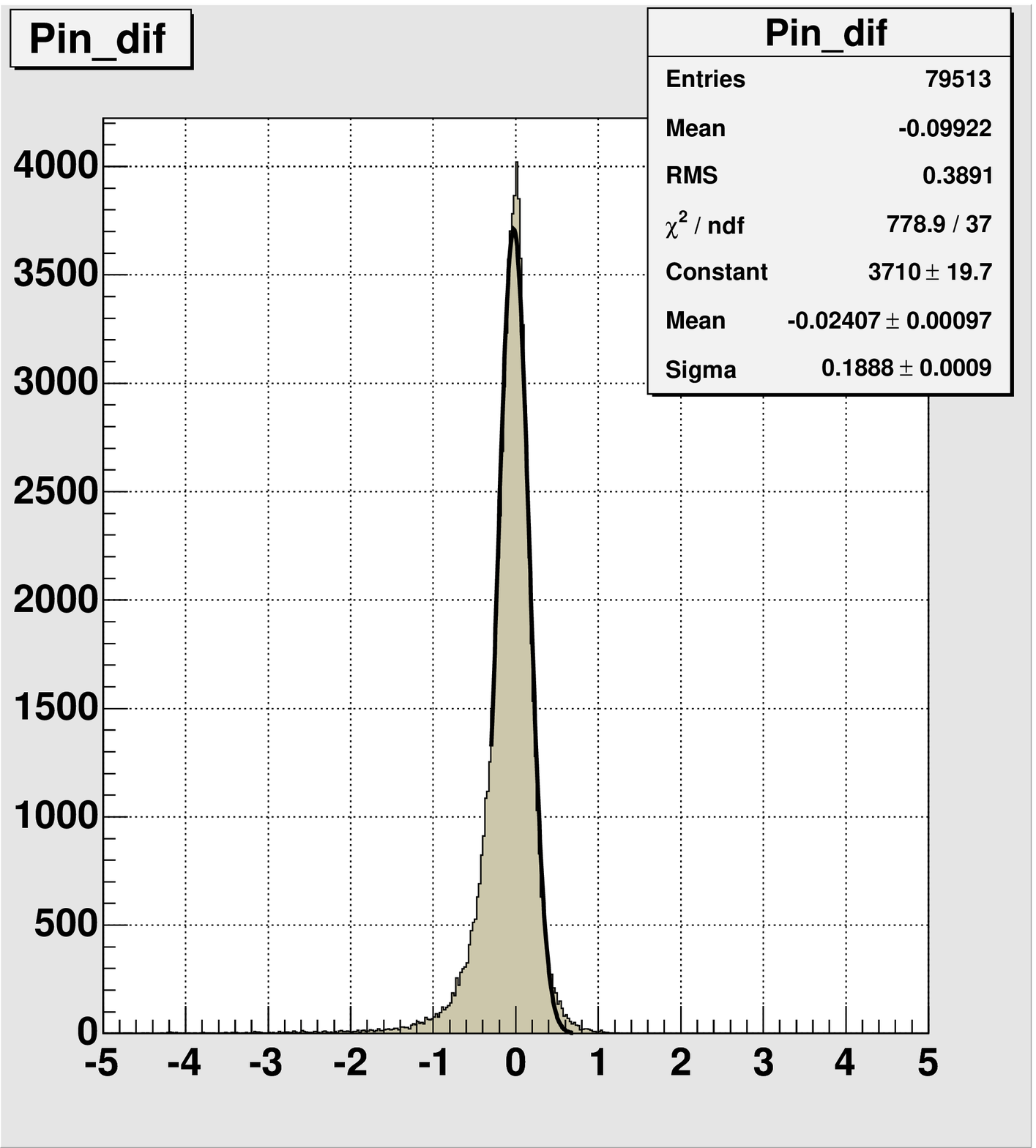}
  \includegraphics[width=0.5\textwidth]{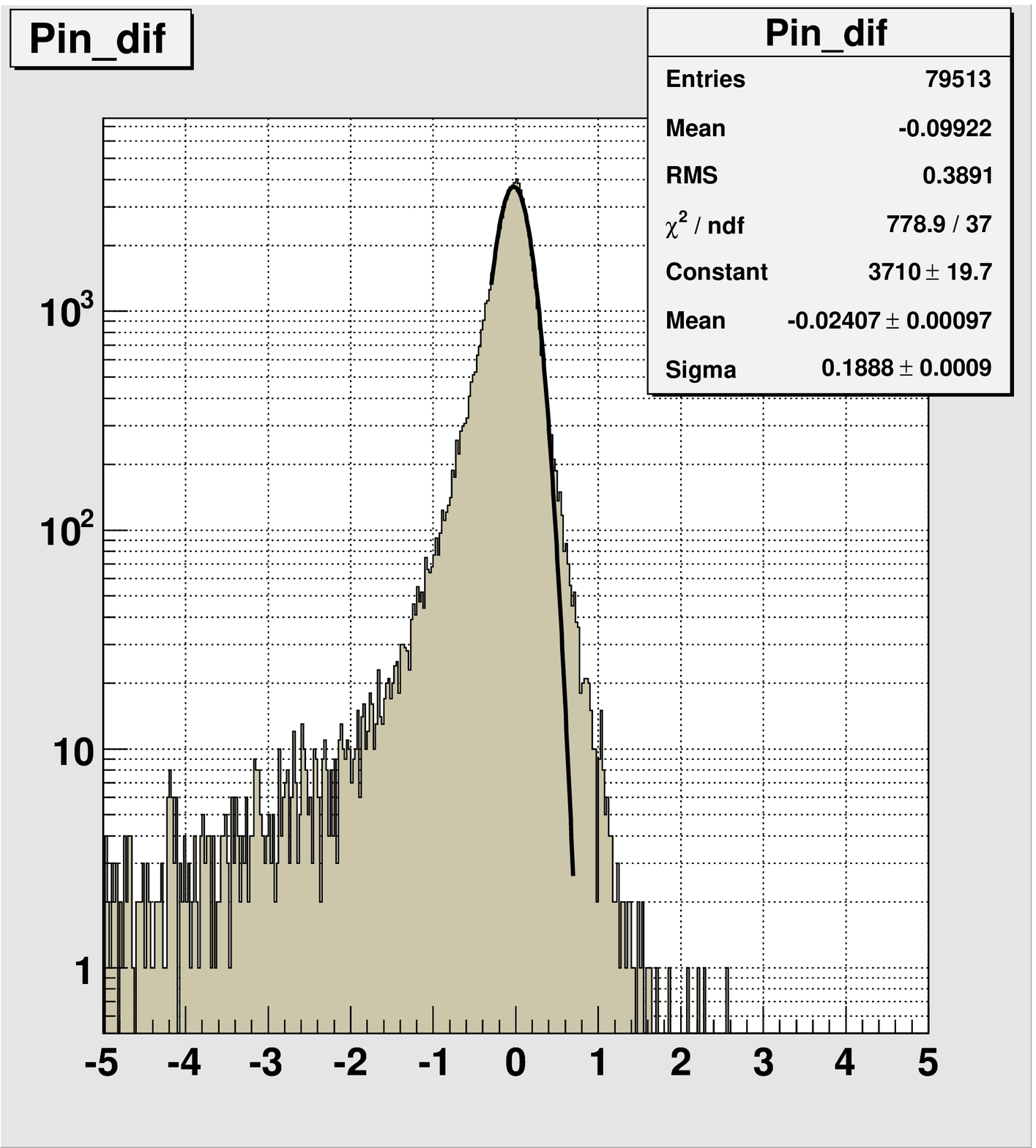} }}
\caption{
 Distribution in the difference between the input reconstructed momentum 
 and the generated input momentum with the background rate of 550 kHz, 
 delta-ray and straw efficiency 97$\%$.
 }
\label{fig:pin_difb1}
\end{figure}

The resolution function obtained as a result of simulation differs from the 
Gaussian form due to non Gaussian tails, which appear due to multiple 
scattering, radiation processes and a non-ideal reconstruction procedure. 
A pattern recognition and track reconstruction procedure based on the Kalman filter
technique developed in \cite{dk_seeds} was applied taking into account backgrounds, 
delta-rays and straw inefficiency.
 The distribution in the difference between the incident momentum ($P_{in}^{f}$) 
reconstructed by
the forward Kalman filter procedure and the generated incident momentum 
($P_{in}$) is shown in
Figure ~\ref{fig:pin_difb1} in linear (left) and logarithmic (right) scale.
According to this distribution the intrinsic tracker resolution is
$\sigma$ = 0.189 MeV/c if one fits the distribution by a Gaussian in the range
-0.3 to 0.7 MeV/c . It follows from Figure ~\ref{fig:pin_difb1}
that outside the range -0.4 to 0.4 MeV the distribution is non Gaussian.

Using this histogram, normalized by the total area of the histogram, as the resolution function,
the probability
to have a signal from muon decay in orbit above a threshold can be presented in
the form

\begin{equation}
P = \sum_i \frac{n_i}{n} \int\limits_{E_{max} + \Delta - \tilde p_i}^{E_{max}} N(E) dE 
\label{eqHistSimp}
\end{equation}

where $n_i$ is the number in $i^{th}$ bin of the histogram, 
$n = \sum_i n_i$ is the total number of events,
$\tilde p_i = (p_i^{min} + p_i^{max})/2$,
$p_0^{min} = \Delta$. 

In the two term approximation Eq.(\ref{eqNappr}) for the electron spectrum 
this equation can be rewritten as

\begin{equation}
P = \sum_i\frac{n_i}{n} \cdot \left[\frac{C_0}{6}(\tilde p_i - \Delta)^6 +
\frac{C_1}{7}(\tilde p_i - \Delta)^7 \right]
\label{eqHistAppr}
\end{equation}
 
\begin{figure}[htb!]
  \centering
   \includegraphics[width=0.7\textwidth]{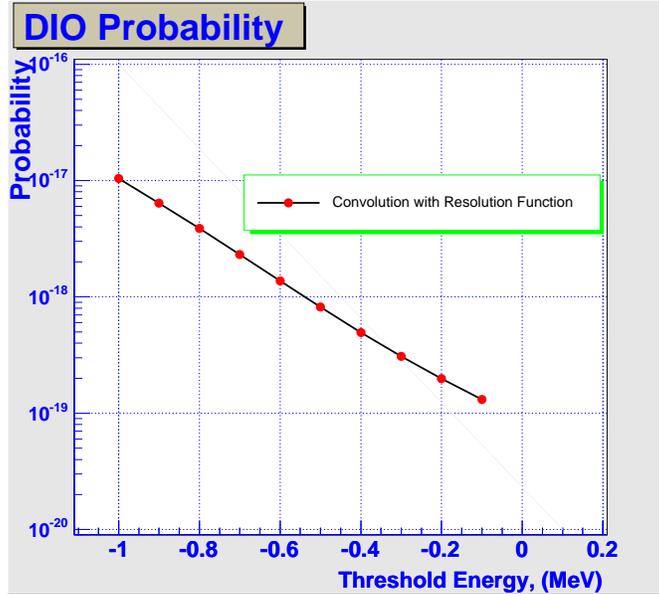}
  \caption{
  Probability of DIO contribution to a signal region as a function of the 
threshold energy $\Delta$ for a resolution
function of the tracker obtained from the simulation.
 }
\label{fig:real_dio}
\end{figure}

Figure ~\ref{fig:real_dio} shows the probability of DIO contribution to 
the signal region as a function of the 
threshold energy $\Delta$ for a resolution
function of the tracker obtained from the simulation (Figure ~\ref{fig:pin_difb1}).  

One can see that in a logarithmic scale the probability decreases approximately
linearly with an increase in threshold energy. This behavior differs from
the sharp decrease of the probability in the case of Gaussian resolution
function and is related to non Gaussian tails.
The existence of non Gaussian tails leads to an increase in probability of
DIO to contribute to the signal region but this additional contribution is
not very significant. For example at the threshold energy of interest 
$\Delta = - 0.7 MeV$ $P = 2.46 \cdot 10^{-18}$ in the case of simulated
resolution function and  $P = 1.93 \cdot 10^{-18}$ in the assumption of
Gaussian resolution function with $\sigma = 0.189 MeV/c$

By using the two term approximation for the electron spectrum and the simulated 
resolution function,
the expected number of primary DIO events in the tracker during the experiment
with an Al target has been calculated and is presented in Table ~\ref{table:events_track}.

\begin{table}[htb!]
\begin{center}
\begin{tabular}{|c|c|c|c|c|c|}
\hline
& & & & & \\
~~~$\Delta$, MeV  ~~~&~~~ -0.9 ~~~&~~~ -0.7 ~~~&~~~ -0.5 ~~~&~~~ -0.3 ~~~&~~~ -0.1 ~~~\\
& & & & & \\
\hline
& & & & & \\
N     &0.68 &0.25 &0.087 &0.03 &0.014 \\
& & & & & \\
\hline
\end{tabular}
\caption { Expected number N of DIO events in a tracker 
in dependence on the threshold energy $\Delta$. Realistic resolution function 
and overall acceptance are included.}
\label{table:events_track}
\end{center}
\end{table}

\section*{DIO events in the calorimeter}

\begin{figure}[htb!]
  \centering
   \includegraphics[width=0.7\textwidth]{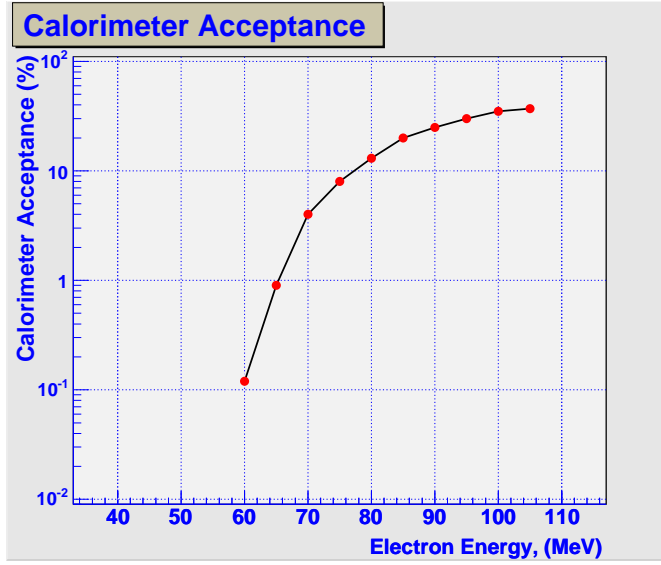}
  \caption{
Geometrical acceptance of the calorimeter.
 }
\label{fig:geom_accept}
\end{figure}

A calorimeter geometrical acceptance 
(Figure ~\ref{fig:geom_accept}) 
plays an important role in suppression of low energy charged particles.

One can see that due to a rapid fall-off
the geometrical acceptance suppresses the
electron spectrum below 70 MeV, reducing significantly in this way
the number of DIO events detected by the
calorimeter above the threshold energy. 

\begin{figure}[htb!]
  \centering
   \includegraphics[width=0.7\textwidth]{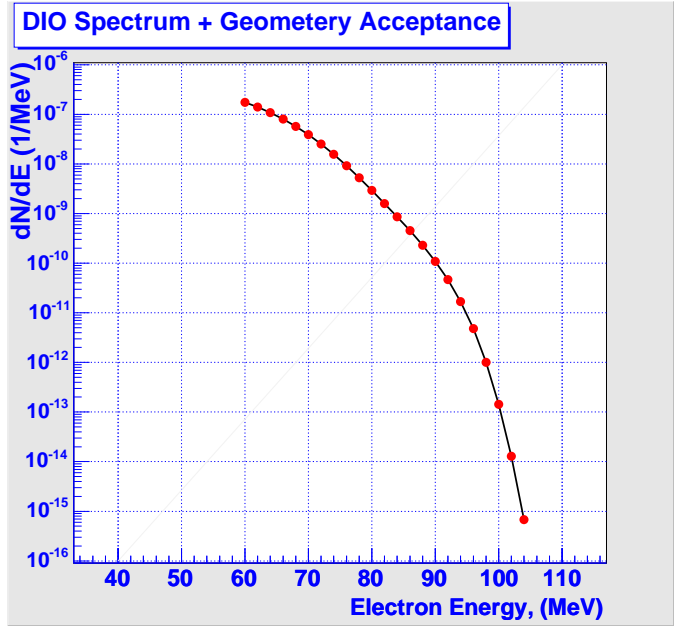}
  \caption{
Differential energy spectrum of
DIO electrons multiplied by the calorimeter geometrical acceptance. This spectrum
is normalized to free muon decay.
 }
\label{fig:geom_cross_spectr}
\end{figure}

Figure ~\ref{fig:geom_cross_spectr} shows a differential energy spectrum of
DIO electrons multiplied by the calorimeter geometrical acceptance.

One can see that in the region below 70 MeV the rapid fall-off of 
the geometrical acceptance with decreasing electron energy almost
compensates the rapid growth of DIO electron spectrum. In particular in
the range from 70 MeV to 60 MeV an increase in the electron spectrum
by 2 orders of magnitude is reduced to a factor 2 due to the fall-off
in the geometrical acceptance.  

The probability P
to have a signal in the calorimeter from muon decay in orbit above a threshold
is given by Eq.(~\ref{eqPfin}) with the electron spectrum multiplied by the 
geometrical acceptance $F_{geom}$:

\begin{equation}
P = \int\limits_{\Delta}^{E_{max} + \Delta} f_G(y) dy \int\limits_{E_{max} + \Delta - y}^{E_{max}} 
N(E) \cdot F_{geom}(E)  dE 
\label{eqPcalo}
\end{equation}

where, as above, $y = E_M - E$, and $E_M$ is the measured electron energy, 
E is the true energy at which an electron was emitted.

\begin{figure}[htb!]
  \centering
   \includegraphics[width=0.7\textwidth]{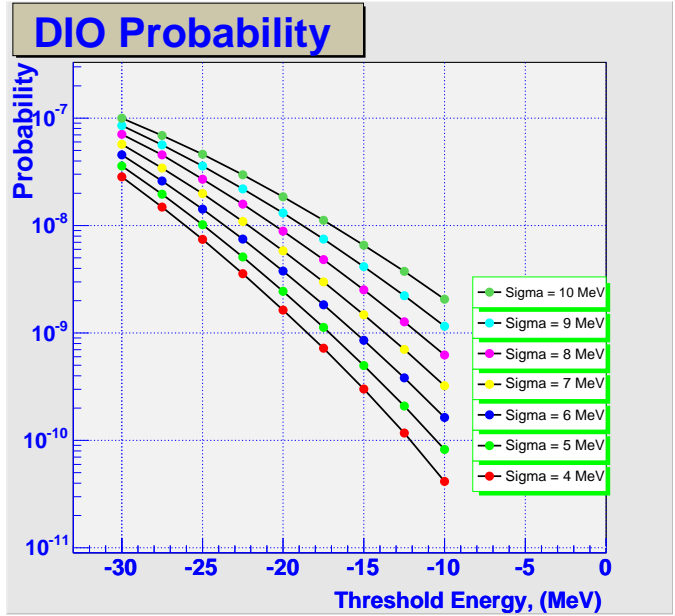}
  \caption{
    Probability of DIO contribution to the calorimeter signal as a function of 
the threshold energy $\Delta$ for different resolutions $\sigma$.
 }
\label{fig:calo}
\end{figure}

In Figure  ~\ref{fig:calo} the probability P of muon decay in orbit contributing 
to the calorimeter signal as a function of
threshold energy $\Delta$ is shown. The resolution function of the calorimeter
is assumed to be Gaussian.

It follows from this plot that the probability is not extremely sensitive to 
the resolution $\sigma$, which is the consequence of fast decreasing
geometrical acceptance below 70 MeV. For example, for 
the measured energy above 80 MeV (the threshold energy
$\Delta = -25 ~ MeV $) even the change in $\sigma$ from 5 MeV/c to 10 MeV/c
increases the probability of DIO contribution only by a factor 5. 
According to this plot the probability P is a fast decreasing function of 
the threshold energy in the considered range because above 55 MeV the electron
spectrum is a steep function of energy.

The expected number N of primary DIO events in calorimeter 
during the time of the experiment can be estimated by Eq.(\ref{eqN1}) with 
$\varepsilon_{acc}=1$.

By using the fit for the electron spectrum described above and the  
Gaussian resolution function,
the expected number of primary DIO events in calorimeter during the experiment
with Al target is calculated and is presented in Table ~\ref{table:events_calo}
for $\sigma = $ 5 MeV/c and 8 MeV/c.

\begin{table}[htb!]
\footnotesize
\begin{center}
\begin{tabular}{|c|c|c|c|c|c|}
\hline
 & & & & &\\
$\Delta$, MeV   &-30 &-25 &-20 &-15 &-10\\
 & & & & &\\
\hline
 & & & & &\\
N ($\sigma =$ 5 MeV)     &$1.8 \times 10^{10}$ &$5 \times 10^9$ &$1.2 \times 10^9$ &$2.5 \times 10^8$ &$4.3 \times 10^7$ \\
 & & & & &\\
\hline
 & & & & &\\
N ($\sigma =$ 8 MeV)     &$3.5 \times 10^{10}$ &$1.4 \times 10^{10}$ & $4.5 \times 10^9$&$1.3 \times 10^9$ &$3 \times 10^8$ \\
 & & & & &\\
\hline
\end{tabular}
\caption { Expected number N of DIO events in calorimeter in dependence on
the threshold energy $\Delta$. Convolution and geometrical acceptance are included.}
\label{table:events_calo}
\end{center}
\end{table}

\section*{Muon conversion}

Muon conversion process is characterized by the appearance of a mono-energetic electron
with energy $E^{max}$ = 104.963 MeV for aluminum.
The rate of $\mu e$ - conversion $R_{\mu e}$ is normalized by the muon capture rate:

\begin{center}
$R_{\mu e} = \Gamma_{\mu e}/ \Gamma_{\mu cap}
$
\end{center}
where $\Gamma_{\mu e}$ is a width for muon conversion on nucleus and $ \Gamma_{\mu cap}$
is a width for muon capture by nucleus.

The total width is defined by

\begin{center}
$\Gamma_{\mu Total} = \Gamma_{\mu free} + \Gamma_{\mu cap}
$
\end{center}

or in terms of life times

\begin{center}
$\frac {1}{\tau_{\mu Total}} = \frac{1}{\tau_{\mu free}} + \frac{1}{\tau_{\mu cap}}
$
\end{center}

where for aluminum $\tau_{\mu Total} = 0.864 \mu sec$ and $\tau_{\mu free} = 2.2 \mu sec$.

The probability $P_{\mu e}$ for muon conversion is expressed as:

\begin{center}
$P_{\mu e} = \Gamma_{\mu e}/ \Gamma_{\mu Total} = R_{\mu e} \cdot \Gamma_{\mu cap}
/ \Gamma_{\mu Total} = 0.6 \cdot R_{\mu e}
$
\end{center}

The probability to have a signal from muon conversion on aluminum is given by:

\begin{equation}
P_{\mu e}^{sig} = P_{\mu e} \cdot \int_{\Delta}^{\infty}{f(y)dy} ~=~
 0.6 \cdot {R_{\mu e}} \cdot \int_{\Delta}^{\infty}{f(y)dy}
\label{Pmue}
\end{equation}

The background to the signal ratio is expressed as:

\begin{center}
$P_{DIO}^{sig}/P_{\mu e}^{sig} = 
0.103 \cdot (10^{-16}/R_{\mu e}) \cdot ~\int_{\Delta}^{\infty}{(y - \Delta)^{6}~f(y)dy}\Big /
 \int_{\Delta}^{\infty}{f(y)dy}
$
\end{center}

where it was used that $C_0/(6 \cdot P_{\mu e}) = 0.103 \cdot 10^{-16}/R_{\mu e}$.

For a Gaussian resolution function this ratio can be approximated by 

\begin{center}
$P_{DIO}^{sig}/P_{\mu e}^{sig} = 
0.103 \cdot (10^{-16}/R_{\mu e}) \cdot \sigma^6 \cdot (15 + 45u^2 + 15u^4 + u^6)
$
\end{center}

where as above $u = \Delta / \sigma$. The precision of this formula is better than $2\%$ 
for u $<$ -2 .

The expected number of registered muon conversion events 
can be calculated from Eqs.(\ref{eqN}),(\ref{Pmue}). For
$R_{\mu e}$ = 10$^{-16}$ sensitivity and setup overall acceptance = 20\%, 
the expected number of muon conversion events is:

\begin{center}
\vspace{0.3cm}
$N = 4\cdot 10^{13} \times 2.5 \cdot 10^{-3} \times 10^{7} \times 0.5
\times 0.2 \times 0.6 \times 10^{-16} = 6.0 $
\end{center}

\section*{Conclusion}

In this memo we studied the detection of electrons from muon decay in
orbit. These electrons are the dominant source of background
for muon - electron conversion experiments because the endpoint of DIO electrons
is the same as the energy of electrons from elastic muon - electron conversion. 

It was found that near the endpoint (E $\simeq $ 100MeV) the  Shanker's 
formula ~\cite{shank}, obtained  
by neglecting the variation of the weak-interaction matrix elements with energy
in DIO process
is in a good agreement
with the results of numerical
calculations  ~\cite{wat} properly taking into account relativistic electron
wave functions and the effect of finite nuclear size on the wave functions. 

It is important to note that in order to simulate a process with 
the threshold in measured energy $E_m^{th}$ one has to take 
into account DIO electrons produced in target starting 
approximately at $E^{th} = E_m^{th} - 2\sigma$. In particular, 
for $E_m^{th} = 80 MeV$ and $\sigma = 5 MeV$ it would correspond
to $E^{th}$= 70 MeV. 
In this case the number of primary DIO events would reach 
$N = 5.65 \times 10^{11}$ making event by event simulation
unfeasible. We assume that present realistic threshold for calorimeter
measured energy is about 90 MeV and the minimal true energy of DIO
electrons is about 80 MeV. This corresponds to the number of DIO
events $N = 1.13 \times 10^{10}$ which is about 50 times less than
in the case of 80 MeV threshold in measured energy.  

The probability of DIO contribution to a signal region was considered
for the tracker with Gaussian resolution function and with the realistic
resolution function obtained in the application of pattern recognition
and momentum reconstruction Kalman filter based procedure to 
GEANT simulated DIO events.

It was found that non Gaussian tails in the simulated 
resolution function do not lead to a significant increase in DIO
contribution to the signal region. The expected number of detected DIO events
during the time of the experiment for realistic resolution function 
was calculated to be about 0.25 if the threshold energy $\Delta = -0.7 ~MeV$
and the overall acceptance is $20 \%$.

It was shown that 
the probability of DIO
contribution is very sensitive to the
threshold energy and proportional to $\Delta^6$ in the limit $|\Delta| \gg \sigma$
where $\sigma$ is the tracker resolution.   

The probability of DIO contribution to the calorimeter signal was studied 
in dependence on the resolution, assuming a Gaussian resolution function
of the calorimeter. In this study the geometrical acceptance 
played an important role,
suppressing DIO contribution of the intermediate range electrons
from muon decay in orbit. 
It was found that the probability of DIO contribution 
is not extremely sensitive to 
the resolution $\sigma$, which is the consequence of the fast decreasing
geometrical acceptance below 70 MeV. For example, for 
the measured energy above 80 MeV (the threshold energy
$\Delta > -25 ~ MeV $) even the change in $\sigma$ from 5 MeV/c to 10 MeV/c
increases the probability of DIO contribution only by a factor 5. 

The expected number N of detected DIO events in calorimeter was estimated 
by using the approximation of the
electron spectrum in the range 55 - 100 MeV. For the measured energy
above 80 MeV and assuming a Gaussian resolution function of the calorimeter
with $\sigma$ = 5 MeV it was found that during the time of the experiment
$N = 5 \times 10^9$.

We wish to thank A.Mincer and P.Nemethy
for fruitful discussions and helpful remarks.

\newpage
\section*{Appendix A \\ Coefficients of Shanker's expansion}

The coefficient $\tilde D$, $\tilde E$, $\tilde F$ of Shanker's expansion 
for different materials can be determined by the 
fitting of numerical values
of $\tilde D$, $\tilde E$, $\tilde F$ 
presented in ~\cite{shank} by 4th power polynomial. 

\begin{figure}[htb!]
  \centering
   \includegraphics[width=0.7\textwidth]{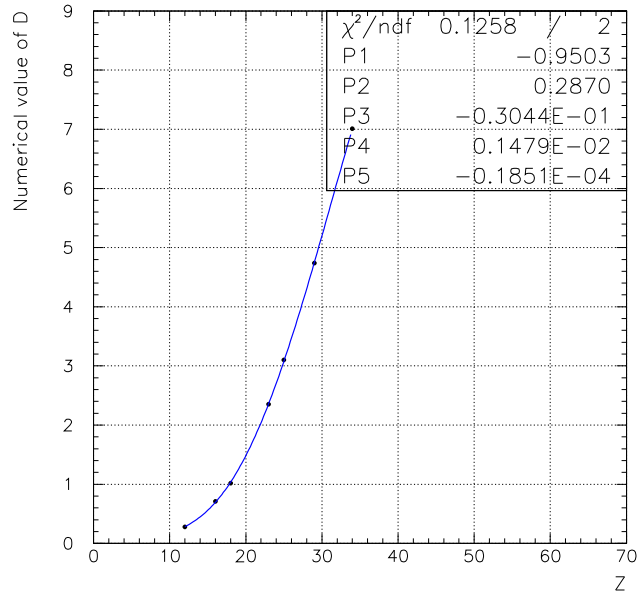} 
  \caption{
  Numerical values of $\tilde D$ coefficient. Solid line is a result of fit by 4th power
  polynomial $f = p1 + p2*x + p3*x^2 + p4*x^3 + p5*x^4$.
 }
\label{fig:D_coef}
\end{figure}

Figure  ~\ref{fig:D_coef} shows the results of the fit for  $\tilde D$
coefficient.

In the same way $\tilde E$ and $\tilde F$ coefficients can be found. 

\begin{figure}[htb!]
  \centering
   \includegraphics[width=0.7\textwidth]{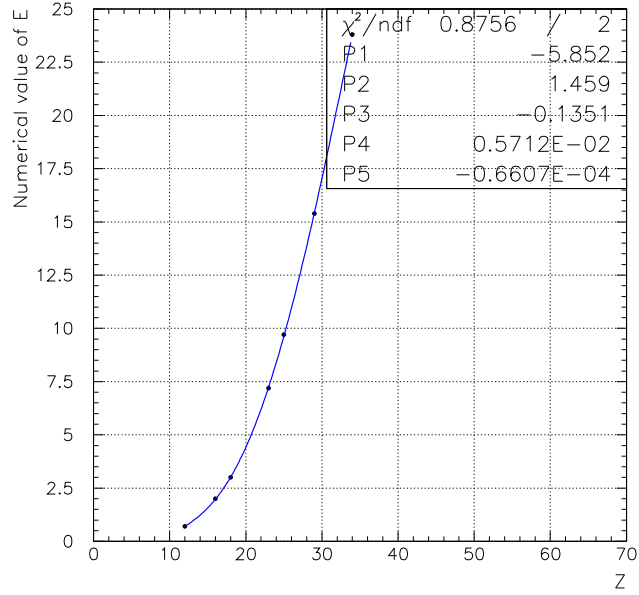} 
  \caption{
  Numerical values of $\tilde E$ coefficient. Solid line is a result of fit by 4th power
  polynomial  $f = p1 + p2*x + p3*x^2 + p4*x^3 + p5*x^4$.
 }
\label{fig:E_coef}
\end{figure}

\begin{figure}[htb!]
  \centering
   \includegraphics[width=0.7\textwidth]{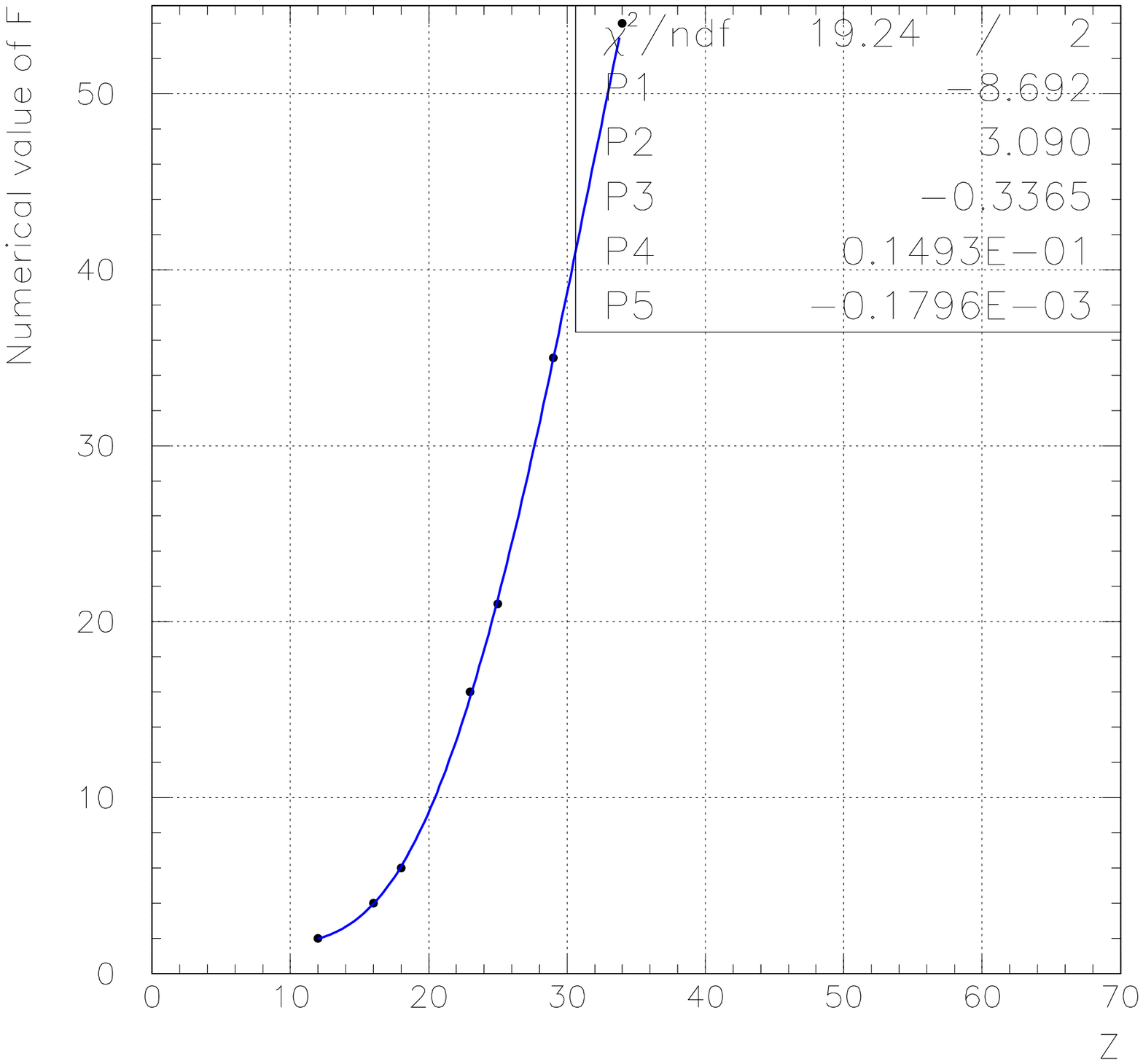}
  \caption{
  Numerical values of $\tilde F$ coefficient. Solid line is a result of fit by 4th power
  polynomial  $f = p1 + p2*x + p3*x^2 + p4*x^3 + p5*x^4$.
 }
\label{fig:F_coef}
\end{figure}

Figures  ~\ref{fig:E_coef} and  ~\ref{fig:F_coef} show the results 
of the fit for D and E coefficients, respectively.

Table ~\ref{table:coef} represents $\tilde D$, $\tilde E$, $\tilde F$ coefficients
for different materials.

\begin{center}
\begin{table}[htb!]
\caption {$\tilde D$, $\tilde E$, $\tilde F$ coefficients for different materials }
\begin{tabular}{|c|c|c|c|}
\hline
  & & & \\
Z   &$\tilde D$ &$\tilde E$ &$\tilde F$ \\
  & & & \\
\hline
~~ 10 ~~&~~~ 0.169990093 ~~~&~~~ 0.281679779 ~~~&~~~ 1.68660045 ~~~\\
\hline
~~ 11 ~~&~~ 0.221416339 ~~&~~ 0.487800866 ~~&~~ 1.8169477 ~~\\
\hline
~~ 12 ~~& 0.282654375& 0.704645038& 1.99828386\\
\hline
~~ 13 ~~& 0.357468009& 0.948250234& 2.27059889\\
\hline
~~ 14 ~~& 0.449176699& 1.2330687& 2.6695714\\
\hline
~~ 15 ~~& 0.560655653& 1.57196677& 3.22657013\\
\hline
~~ 16 ~~& 0.694335938& 1.97622538& 3.96865273\\
\hline
~~ 17 ~~& 0.852204204& 2.45553946& 4.91856623\\
\hline
~~ 18 ~~& 1.03580284& 3.01801825& 6.09474707\\
\hline
~~ 19 ~~& 1.24623013& 3.67018533& 7.51132059\\
\hline
~~ 20 ~~& 1.48413992& 4.41697836& 9.17810249\\
\hline
~~ 21 ~~& 1.74974191& 5.26174974& 11.1005974\\
\hline
~~ 22 ~~& 2.04280138& 6.20626593& 13.2799997\\
\hline
~~ 23 ~~& 2.36263967& 7.25070715& 15.713191\\
\hline
~~ 24 ~~& 2.70813346& 8.39366913& 18.3927441\\
\hline
\end{tabular}
\label{table:coef}
\end{table}
\end{center}

\section*{Appendix B \\ Fit of electron spectrum}

Below 55 MeV the electron spectrum $N_{free}(E)$ can be described by an 8 parameter fit 
in the form $e^{f(E)}$ where f(E) is a polynomial of power 7:

\begin{equation}
f(E) = f_7 \cdot E^7 + f_6 \cdot E^6 + f_5 \cdot E^5 + f_4 \cdot E^4 +
f_3 \cdot E^3 + f_2 \cdot E^2 + f_1 \cdot E + f_0 
\end{equation}
 
For aluminum the fit, using numerical results for the electron spectrum
~\cite{wat}, gives: 

\begin{center}
$f_7 = -4.1321585902527747 \times 10^{-10}$; \\
$f_6 = 7.26144332670667 \times 10^{-8}$; \\
$f_5 = -5.0391208057229 \times 10^{-6}$; \\
$f_4 = 1.720029756203247 \times 10^{-4}$; \\ 
$f_3 = -2.797753809879399 \times 10^{-3}$; \\
$f_2 = 1.003520948296598 \times 10^{-2}$; \\
$f_1 = 0.35027124751754113$; \\
$f_0 = -8.495230158021982$; \\
\end{center}

Above 55 MeV the electron spectrum $N_{free}(E)$ can be described by an 8 parameter fit 
in the form $e^{g(E)}$ where g(E) is a polynomial of power 7:

\begin{equation}
f(E) = g_7 \cdot E^7 + g_6 \cdot E^6 + g_5 \cdot E^5 + g_4 \cdot E^4 +
g_3 \cdot E^3 + g_2 \cdot E^2 + g_1 \cdot E + g_0 
\end{equation}
 
For aluminum the fit, using numerical results for the electron spectrum
~\cite{wat}, gives: 

\begin{center}
$g_7 = 5.882860554682616 \times 10^{-10}$; \\
$g_6 = -3.311494207831439  \times 10^{-7}$; \\
$g_5 = 7.90482092857202 \times 10^{-5}$; \\
$g_4 = -1.0375709354278761 \times 10^{-2}$; \\
$g_3 = 0.808726145419064$; \\
$g_2 = -37.4179819613308$; \\
$g_1 = 950.4758914390899$; \\
$g_0 = -10212.91359983421$; \\
\end{center}

\end{document}